\documentclass[12pt]{iopart}
\usepackage[dvips]{graphicx}
\usepackage[english]{babel}
\usepackage{verbatim}
\usepackage{iopams}  
\usepackage{amsfonts}
\usepackage{makeidx}
\usepackage{color}
\usepackage{subfigure}
\usepackage{multirow}
\newcommand{\pp}{\rm pp}
\newcommand{\sqrts}{\sqrt{s}}
\newcommand{\sqrtsNN}{\sqrt{s}_{\rm NN}}
\newcommand{\gsim}{\,{\buildrel > \over {_\sim}}\,}

\newcommand{\ctau}{{\rm c}\tau}

\newcommand{\GeV}{\mathrm{GeV}}
\newcommand{\TeV}{\mathrm{TeV}}

\newcommand{\gev}{\mathrm{GeV}}
\newcommand{\tev}{\mathrm{TeV}}

\newcommand{\mum}{\mathrm{\mu m}}

\newcommand{\ptrans}{p_{\rm t}}

\newcommand{\DtoKpi}{{\rm D^0\to K^-\pi^+}}
\newcommand{\DtoKpipi}{{\rm D^+\to K^-\pi^+\pi^+}}

\newcommand{\Dzero}{{\rm D^0}}

\newcommand{\Dplus}{{\rm D^+}}

\newcommand{\RAA}{R_{\rm AA}}
\newcommand{\RAAD}{R_{\rm AA}^{\rm D}}
\newcommand{\RAAB}{R_{\rm AA}^{\rm B}}

\newcommand{\avNcoll}{\langle N_{\rm coll} \rangle }
\newcommand{\avTAA}{\langle T_{\rm AA} \rangle}

\begin{document}

\title[]{{D meson nuclear modification factors in Pb--Pb collisions
at $\sqrtsNN = 2.76~\TeV$, measured with the ALICE detector}}

\author{A~Rossi for the ALICE Collaboration}

\address{Universit\`a di Padova and INFN - Sezione di Padova, Padova, Italy}
\ead{rossia@pd.infn.it}
\begin{abstract}
  The ALICE experiment has measured the D meson 
  production in pp and Pb-Pb collisions at the LHC at $\sqrt{s}=7$ and $2.76~{\rm TeV}$ 
  and $\sqrt{s_{NN}} = 2.76~{\rm TeV}$ respectively, via the exclusive reconstruction of hadronic 
  decay channels.  
  The analyses of the 
  ${\rm D^{0} \rightarrow K^{-}\pi^{+}}$ and ${\rm D^{+} \rightarrow K^{-}\pi^{+}\pi^{+}}$ 
  channels will be described and the preliminary results for the ${\rm D^{0}}$ and ${\rm D^{+}}$ 
  nuclear modification factor will be presented.
\end{abstract}
%
\section{Introduction}
%
The comparison of heavy flavour production in proton-proton and heavy-ion collisions allows 
to probe the properties of the high-density QCD medium formed in the latter and to study the 
mechanism of in-medium partonic energy loss~\cite{Dokshitzer:2001zm}. 
A sensitive observable is the nuclear modification factor,
defined, for a particle species $h$, 
as $\RAA^{h}(\ptrans)=\frac{{\rm d}N^{h}_{\rm AA}/{\rm d}\ptrans}{\avTAA\times {\rm d}\sigma^{h}_{\pp}/{\rm d}\ptrans}$,
where $N^{h}_{\rm AA}$ is the 
normalized yield measured
in heavy-ion collisions, $\avTAA$ is the average nuclear overlap function calculated with the Glauber model in the
considered centrality range, and $\sigma^{h}_{\pp}$ is the production cross-section 
in pp collisions. 
For hard processes, in the absence of any medium and cold nuclear matter effect, $\RAA=1$ is expected.
By comparing the nuclear modification factors 
of charged pions $(\RAA^{\pi^{\pm}})$, mostly originating
from gluon fragmentation, with that of hadrons with charm $(\RAAD)$ and beauty $(\RAAB)$
the dependence of the energy loss on the parton nature (quark/gluon) and mass can be investigated.
A mass ordering pattern $\RAA^{\pi^{\pm}}(\ptrans)\lesssim \RAAD(\ptrans)\lesssim \RAAB(\ptrans)$ has been predicted~\cite{Armesto:2005iq}.
In these proceedings the measurement of the $\Dzero$ and $\Dplus$ $\RAA$ 
in Pb--Pb collisions at the LHC is presented. More information on the ALICE detector can be found
in~\cite{aliceJINST,JurgenProcQM}; the procedure to determine the collision centrality
is described in~\cite{AlbericaProcQM}.

\section{Measurement of D meson production with the ALICE detector} 

\begin{figure}[!t]
  \begin{center}
    \includegraphics[height=0.27\textheight]{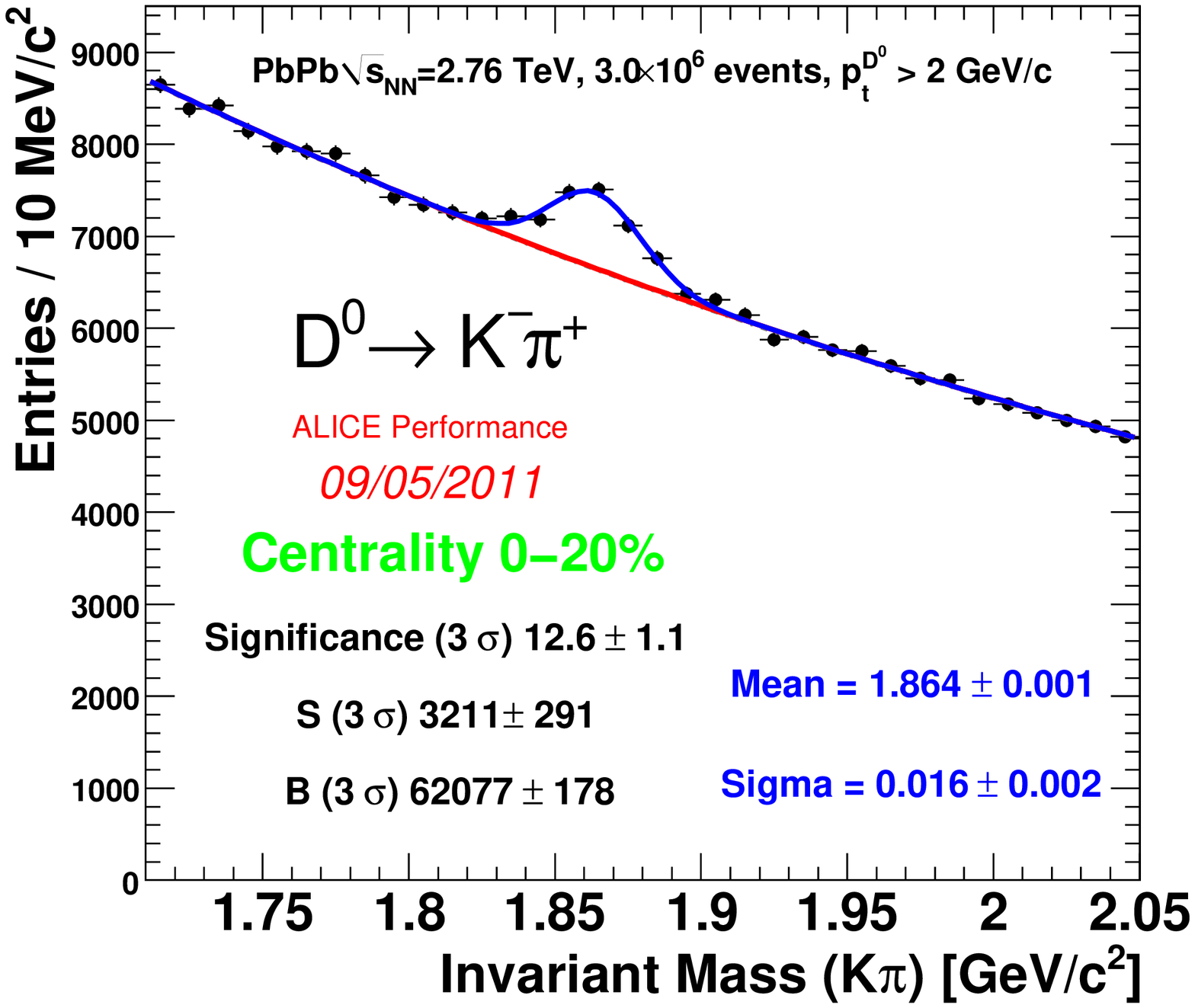}
    \includegraphics[height=0.26\textheight]{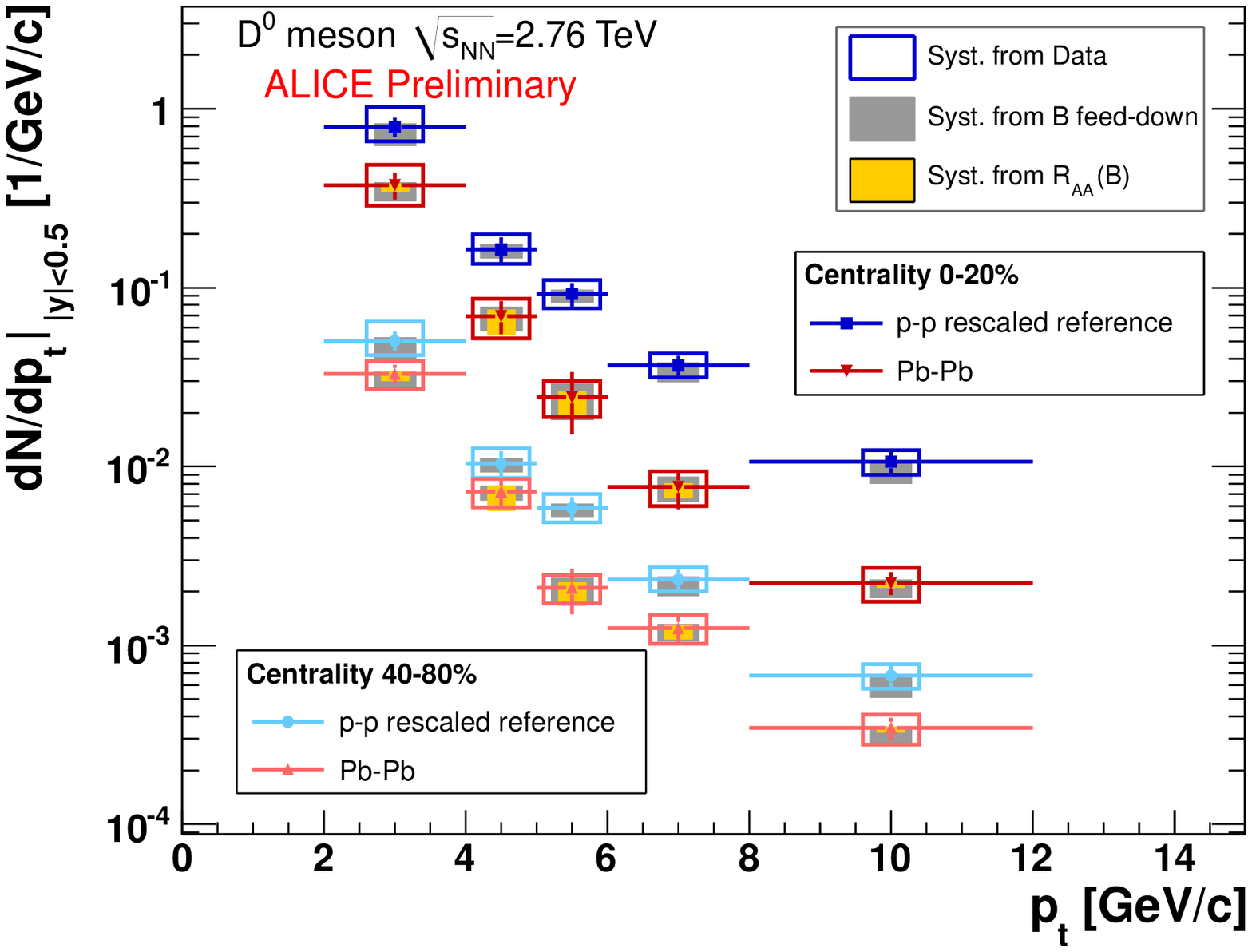}
    \caption[]{Left: invariant mass distributions of $\Dzero$ candidates 
      for $\ptrans>2~\gev/c$ 
      obtained from the analysis
      of $\sim 3\times 10^6$ central ($0-20\%$) Pb--Pb events at $\sqrtsNN=2.76~\tev$. 
      The curves show the fit functions as described in the text.
      Right: $\Dzero$ spectra for central ($0-20\%$) and peripheral ($40-80\%$) Pb--Pb events at $\sqrtsNN=2.76~\tev$  
      compared to the reference spectra obtained by scaling the ALICE measurement in $\pp$ collisions at $7~\tev$ (see text).
    }
    \label{fig:invmassCentSpectra} 
  \end{center}
\end{figure}

The production of $\Dzero$ and $\Dplus$ ($\ctau\approx 123$ and $312~\mum$ respectively~\cite{pdg}) 
was measured in pp and Pb--Pb collisions at central rapidity ($|y|<0.5$) via the exclusive reconstruction 
of the decays $\DtoKpi$ (with branching ratio, BR=$3.89\pm 0.05\%$~\cite{pdg}) 
and $\DtoKpipi$ (BR=$9.4\pm 0.4\%$~\cite{pdg}).
The analysis strategy for the extraction of the $\Dzero$ and $\Dplus$ signals
out of the large combinatorial background from uncorrelated tracks 
is based on the reconstruction 
and selection of secondary vertex topologies with significant separation (typically a few hundred micrometer) 
from the primary vertex. 
The Time Projection Chamber (TPC) and the Inner Tracking System (ITS) detectors provide a spatial resolution
on the track position in the vicinity of the primary vertex of the order of few tens of microns
at sufficiently high $\ptrans$~\cite{RossiVertex2010}.
Tracks displaced from the primary vertex are selected to reconstruct D meson candidates.
Good alignment between the reconstructed
meson momentum and the flight direction between the primary and secondary
vertex is also required.
The identification of the charged kaon in 
the TPC and Time Of Flight detectors provides additional background rejection. 
A particle identification strategy
that preserves most of the D meson signal was adopted.
Similar analyses were performed on $\pp$ and Pb--Pb data, with
a tighter selection in the latter case, dictated by the higher combinatorial
background.
To extract the signal, a fit to the invariant mass distribution is performed,
using a Gaussian function for the signal peak and an exponential function for the 
background. Figure~\ref{fig:invmassCentSpectra} (left) shows the invariant mass
distribution of $\Dzero$ candidates for $\ptrans>2~\gev/c$, obtained from the analysis
of $\approx 3 \times 10^6$ Pb--Pb events in the $0-20\%$ centrality range. 
The `raw' signal is corrected for acceptance and
efficiency using Monte Carlo simulations based on Pythia (Perugia-0 tuning) and HIJING event generators.
%
The contribution of D mesons from B decays was evaluated
relying on the FONLL prediction, which 
describes well bottom production at the Tevatron~\cite{fonllBcdf} and at the 
LHC~\cite{lhcbBeauty,cmsJpsi}. 
In the Pb--Pb case, 
the FONLL prediction of secondary D mesons production in pp collisions
is multiplied by $\avNcoll \times \RAAB$, where
the hypothesis on the B meson $\RAA$ encodes all
potential nuclear and medium effects affecting B production. 
The systematic error due to the FONLL theoretical uncertainty on 
beauty prediction was estimated from the spread 
of the results recovered 
using the minimum and maximum predictions for secondary D production. 
%
%
As reported in~\cite{DaineseQM2011}, the D meson production cross-sections  
in $\pp$ collisions at $\sqrts=7~\tev$ 
are well described by predictions based on 
pQCD calculations as FONLL~\cite{fonllBcdf} and GM-VFNS~\cite{VFNSalice}.
To provide the reference cross-section at $2.76~\tev$, needed to compute the $\RAA$,  
the measurement at $7~\tev$ is scaled by the ratio of FONLL predictions for D meson 
production at $\sqrts=2.76$ and $7~\tev$~\cite{Note276}. 
The uncertainty on the scaling decreases with $\ptrans$ from $25\%$ to $10\%$, 
as estimated by varying the parameters entering the FONLL calculations,
namely the factorization and normalization scales and the 
charm quark mass.
In March 2011 a sample of $\approx 6.5\times 10^{7}$ events from
$\pp$ collisions at $\sqrts=2.76~\TeV$ was collected. 
The $\Dzero$ and $\Dplus$ signals were measured
in the $2<\ptrans<8~\gev/c$ range.
While the accumulated statistics did not 
allow for determining a pp reference over the whole momentum range, 
it served as an important cross check of the theoretical scaling procedure 
in the momentum range where the data sets overlap.
Considering only the statistical errors, 
the ratio between the cross-section measured at $2.76~\tev$
and that obtained from the scaling of the measurement at $7~\tev$
is compatible with 1 within $1\sigma$ 
for the $\Dplus$ while, for the $\Dzero$,
within $1\sigma$ above $4~\gev/c$ and within $2\sigma$ for $2<\ptrans<4~\gev/c$.
A further $20\%$ asymmetric contribution to the systematic error 
in the $2<\ptrans<4~\gev/c$ range has been added 
on the $\Dzero$ cross-section at $2.76~\tev$ used as the $\pp$
reference in the $\RAA$ calculations.
Figure~\ref{fig:invmassCentSpectra} (right) shows the 
$\Dzero$ $\ptrans$ spectra measured in central ($0-20\%$) and peripheral ($40-80\%$) Pb--Pb 
collisions, compared to their respective reference spectra. 
\section{D meson nuclear modification factors}
\begin{figure}[!t]
  \begin{center}
    \includegraphics[height=0.23\textheight]{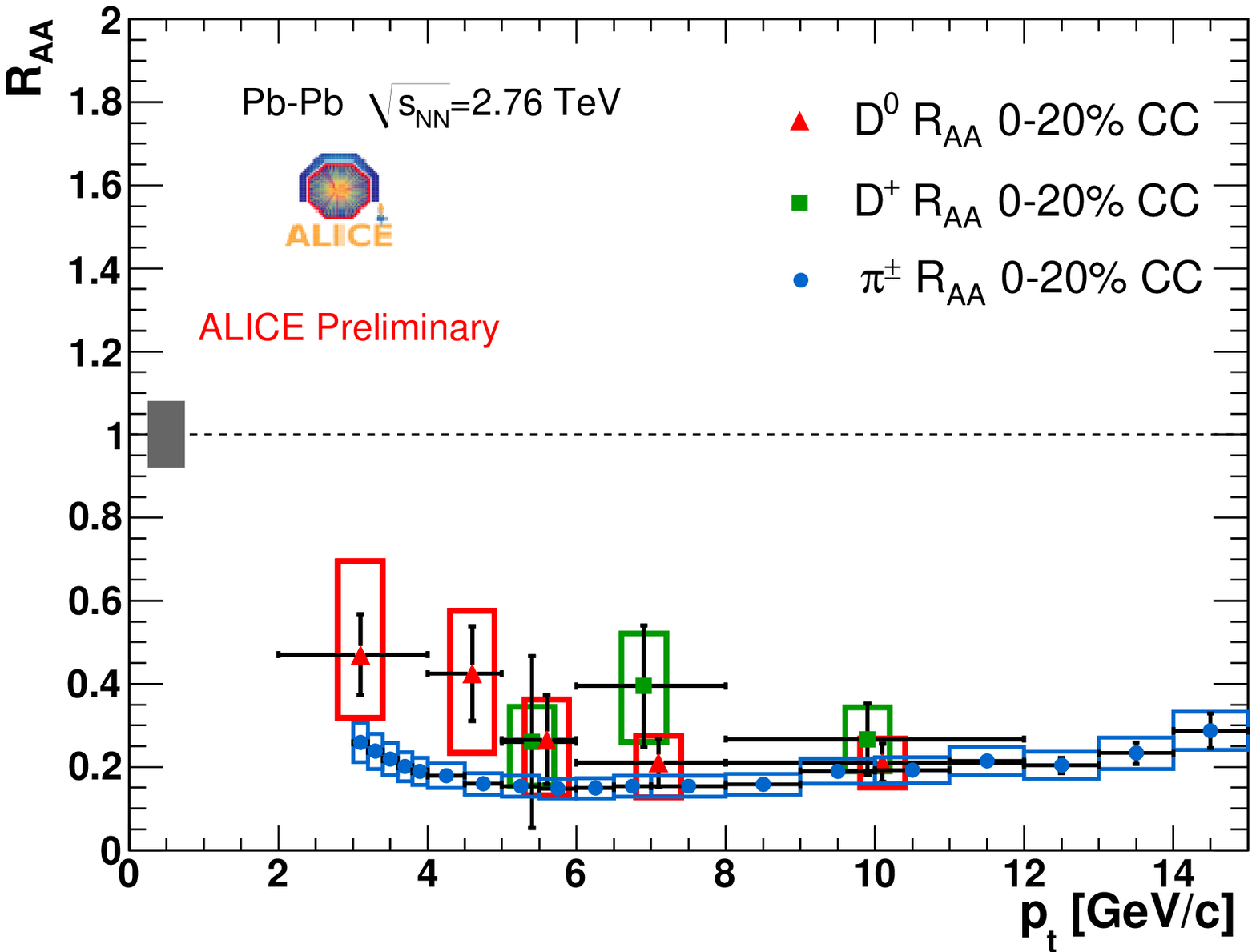}
    \includegraphics[height=0.23\textheight]{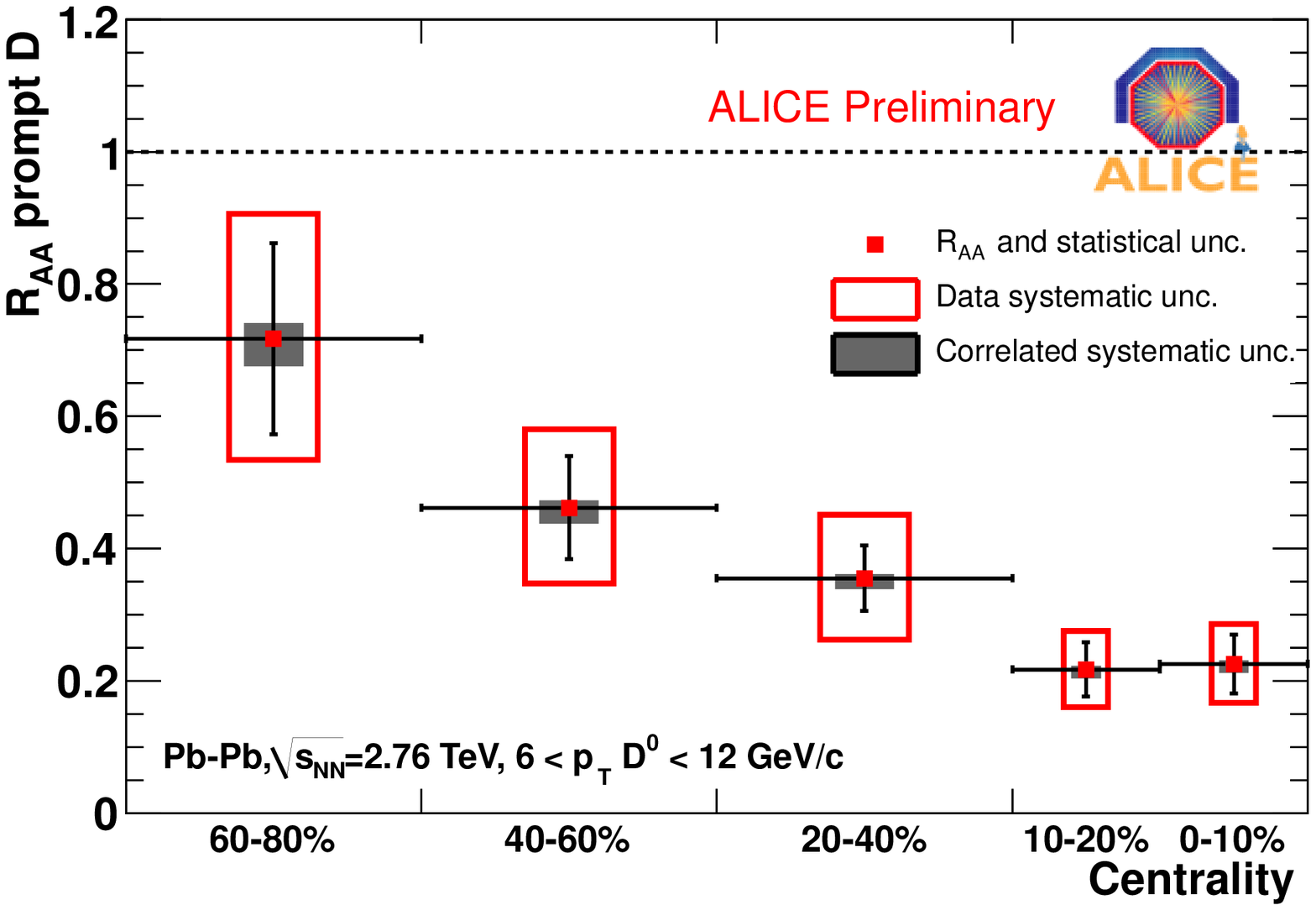}
    \caption[]{Left: $\Dzero$ and $\Dplus$ nuclear modification factor measured in central  ($0-20\%$) 
      Pb--Pb collisions. The vertical lines represent the statistical errors.
      The filled grey box around $\RAA=1$ represents the uncertainty on the 
      normalization of the $\pp$ cross-section and on $\avTAA$. 
      The empty boxes represent the total systematic uncertainties.       
      The charged pions $\RAA$ is shown for comparison.
      Right: $\Dzero$ nuclear modification factor as a function of centrality for $6<\ptrans^{\Dzero}<12~\gev/c$. 
    }
    \label{fig:D0DplusChargedRAA_RAAVSCentrality}    
  \end{center}
\end{figure}

The D meson production is suppressed by a factor 4-5 in central events for $\ptrans\gsim 5~\gev/c$, 
as quantified by the nuclear modification factors shown 
in Fig.~\ref{fig:D0DplusChargedRAA_RAAVSCentrality} (left). The $\Dzero$ and $\Dplus$ $\RAA$
agree within errors.
For the $\Dzero$, the statistical uncertainty is of the order of $20-25\%$, 
the estimated total systematic uncertainty,
accounting for the uncertainties on the signal extraction procedure, on track reconstruction
efficiency, on the MC corrections for reconstruction
acceptance, cut and PID  selection, and for possible differences in the $\Dzero$ and $\overline{\Dzero}$
reconstruction varies from ${}^{+55\%}_{-26\%}$ for $2<\ptrans<4~\gev/c$ to $\pm 25\%$ for $8<\ptrans<12~\gev/c$.
The hypothesis on $\RAAB$ is varied in order to span the range $0.3<\RAAD/\RAAB<3$, with the 
$\RAAD/\RAAB$ calculated a posteriori. 
The range of $\RAAD$ values obtained in each $\ptrans$ bin
is considered as the systematic error due to the $\RAAB$ assumption.
For the $\Dzero$, the maximum spread is of the order of $15\%$ in the bin $2<\ptrans<4~\gev/c$,
$25\%$ in the bin $4<\ptrans<5~\gev/c$ and $10\%$ in the bin $8<\ptrans<12~\gev/c$.
%

In central collisions, $\RAA$ decreases with $\ptrans$ from $\sim 0.45$ at low $\ptrans$
to $\approx 0.2$ at high $\ptrans$. The D meson $\RAA$ is compatible
within errors with the charged pion $\RAA$: 
the larger statistics expected from the higher luminosity Pb--Pb 2011 run and a reduction of the systematic
errors should allow to conclude whether $\RAAD>\RAA^{\pi^{\pm}}$ for $\ptrans\lesssim 5~\GeV/c$ 
and quantify the difference.  
The dependence on the centrality is shown in Fig.~\ref{fig:D0DplusChargedRAA_RAAVSCentrality} (right) 
for $6<\ptrans<12~\gev/c$: $\RAA$ decreases from $\approx 0.7$ in peripheral ($60-80\%$)
to $\sim 0.2$ in central ($0-10\%$) events.
As verified by calculating the $\RAA$ expected from pQCD 
calculation based on the MNR code~\cite{MNR} and nuclear PDF from EPS09 parametrization~\cite{EPS09},
nuclear shadowing yields a relatively small 
effect for $\ptrans\gsim 5~\gev/c$ ($0.85<\RAAD<1.08$ at $\ptrans=5~\gev/c$). 
Therefore, the observed suppression can be considered 
as an evidence of in-medium charm quark 
energy loss. 

\end{document}